# Micromechanical Origin of Heat Transfer to Granular Flow

Xintong Zhang[*,1], Sarath Adapa[*,1], Tianshi Feng[1], Jian Zeng[1], Ka Man Chung[2], Clifford Ho[3], Kevin Albrecht[3], Renkun Chen[#1,2]

[1]Department of Mechanical and Aerospace Engineering, University of California, San Diego, La Jolla, California 92093, United States

[2]Program in Materials Science and Engineering, University of California, San Diego, La Jolla, California 92093, United States

[3]Concentrating Solar Technologies Department, Sandia National Laboratories, 1515 Eubank Blvd. SE, Albuquerque, New Mexico, 87123, United States

[*]These authors contributed equally to this work

[#]Corresponding Author: rkchen@ucsd.edu

**Abstract**

Heat transfer across a granular flow is comprised of two resistances in series: near the wall and within the bulk particle bed, neither of which is well understood due to the lack of experimental probes to separate their respective contribution. Here, we use a frequency modulated photothermal technique to separately quantify the thermal resistances in the near-wall and the bulk bed regions of particles in flowing states. Compared to the stationary state, the flowing leads to a higher near-wall resistance and a lower thermal conductivity of bulk beds. Coupled with discrete element method simulation, we show that the near-wall resistance can be explained by particle diffusion in granular flows.

Dense granular flows are widely used as heat transfer media in particle heat exchangers [1,2], thermal energy storage [3], thermochemical and nuclear reactors [4,5], material processing [6], and catalytic beds [7,8]. Previous analyses of dense granular flows in vertical channels have established that the flow is plug-like in the bulk and has a wall-adjacent shear layer with thickness of 1-10 particle diameters ($D_p$) depending on the wall roughness [9,10]. While the bulk rheology of such flows has been widely studied [11], the near-wall region has seldom been quantified. The heat transfer from a wall to the bulk is critical and sensitive to particle packing structures. In a randomly packed particle bed, the presence of a container wall leads to larger void space near the wall. When the granular medium starts to flow, the shear induced by wall friction can cause further dilation near the interface [12], which has a non-negligible impact on heat transfer in the particle bed [13]. The importance of understanding the near-wall region is reflected in the particle-wall heat transfer calculations which approximate granular flow as a plug-like continuum with a bulk effective thermal conductivity. In the pioneering study by Sullivan and Sabersky [14], they found that a discrepancy from the continuum assumption could be accounted for by a near-wall thermal resistance ($R_{NW}$) of a granular flow. Based on a model fitting of their heat transfer coefficient (HTC) measurements, they attributed this to the presence of an effective air gap layer close to the wall with a thickness of $0.085D_p$. Later, experimental works [15-17] confirmed that only after including the $R_{NW}$ can the measured results be fitted by analytical Nusselt number solutions or numerical models. However, previous measurements on moving particle bed heat transfer failed to directly isolate the $R_{NW}$ from the resistance in the bulk region of the flow. By only monitoring temperature difference along the flow, these measurements lack the spatial resolution needed to separate the near-wall and the bulk regions, both of which could be conceivably dependent on flow velocity. Therefore, the physical understanding of particle-wall heat transfer in granular flows remains elusive.

Efforts have also been made to depict the $R_{NW}$ theoretically. Natarajan and Hunt [18] estimated it using a dense-gas kinetic theory. Surprisingly, their results depicted a lower $R_{NW}$ in flowing particles, which is inconsistent with their previous experiment [19]. Due to collisions and frictions with the wall, particle motion and packing density experience significant change across the thin near-wall layer in granular flows [20-22], resulting in the failure of purely analytical approaches based on the continuum assumption. Recently, discrete element method (DEM) simulation has been utilized to analyze the $R_{NW}$ [6,23,24], but a microscopic mechanism remains unknown. In general, prior experimental and modeling work has yet to provide a clear picture of heat transfer physics in dense granular flows.

In this Letter, we devised a frequency-domain modulated photothermal radiometry (MPR) measurement technique, extended from our earlier work on bulk solids and liquids [25,26], to separately quantify the near-wall thermal resistance and the bulk effective thermal conductivity ($k_{eff}$) of gravity-driven dense granular flows. Since the rigid particles normally have high elastic modulus and little deformation during their contact with the wall, the particle-wall contact area is negligible in terms of heat transfer [27,28]. Therefore, the $R_{NW}$ can be solely attributed to an effective air gap adjacent to the wall with a thickness of $D_{air}$ by $R_{NW} = D_{air}/k_{air}$, where $k_{air}$ is the thermal conductivity of air. When particle beds start flowing, an increasing $D_{air}$ and a decreasing $k_{eff}$ were observed compared to their stationary states. DEM simulations were conducted to acquire particle packing information from dense granular flow confined in a channel, which was later imported into COMSOL Multiphysics® to obtain $k_{eff}$. Besides, we estimated $D_{air}$ using dense-gas kinetic theory based on particle velocity fluctuation from DEM simulations. Both $k_{eff}$ and $D_{air}$ from our modeling agree well with the experiments.

In our experiments, we measured flowing ceramic particles with a mean diameter of 275 μm (CARBOBEAD CP 40/100, hereinafter referred to as CP 40/100) and 404 μm (CARBOHSP 40/70, hereinafter referred to as HSP 40/70) between 300 ºC and 650 ºC (see Supplemental Material [29], S4). Both types of particles are spheroid with an average roundness of around 0.8 and have a bulk-material thermal conductivity between 4.51 W m$^{-2}$ K$^{-1}$ to 5.21 W m$^{-2}$ K$^{-1}$ within the temperature range of this study [3]. The flowing channel setup shown in Fig. 1(b) consists of a hot particle reservoir, a 30-cm-long smooth Inconel 625 channel, and a slide gate at the bottom outlet. The granular flow was confined in the rectangular channel with depth of 5 mm and width of 30 mm. Flow velocity was controlled in the range from 0 mm s$^{-1}$ to 15 mm s$^{-1}$ by changing the slide gate opening. For stationary bed measurements with no need of continuous particle supply, the particles were closely packed in the cavity of an Inconel 625 holder which was heated up by insertion heaters (see Supplemental Material [29], S1). A continuous-wave laser with its intensity modulated at angular frequency $\omega$ was shined on the front side of the channel (a 100 μm stainless steel sheet coated with Pyromark 2500 black paint for light adsorption). The steel sheet is in contact with the particles and conducts the thermal wave to the granular flow. The temperature oscillation amplitude $|\theta_s|$ of the steel sheet was measured by detecting the infrared signal emitted from the black coating using infrared detectors [Fig. 1(a)]. The laser heat flux and thermometry were calibrated by measuring a standard sample of borosilicate glass with known thermal conductivity (see Supplemental Material [29], S3). As shown in Fig. 1(c), the thermal wave at angular frequency $\omega$ can probe into different depths of the sample given by

$$L_p = \sqrt{\frac{2\alpha}{\omega}} \qquad (1)$$

where $\alpha$ is the material thermal diffusivity. By modulating the laser frequency, the MPR technique provides a convenient approach to experimentally quantify the $R_{NW}$ (i.e., the $D_{air}$) and the $k_{eff}$. Fig. 1(d) shows the measured $|\theta_s|$ vs. $1/\sqrt{\omega}$ of HSP 40/70 at 300 ºC under various flow velocities ($U$). $|\theta_s|$ increases with $1/\sqrt{\omega}$, i.e., the penetration depth, and a larger slope of the curve reflects a higher local thermal resistance. We developed a two-dimensional COMSOL Multiphysics® model to describe the continuum plug flow of particles by solving an advection-diffusion equation (see Supplemental Material [29], S5) with parameters from Refs. [25,30-33]. The air gap layer is modelled as a stagnant material adjacent to the granular media and an incident oscillating heat flux perpendicular to the flow is imposed as the lateral boundary condition. The resulting surface temperature oscillations are collected at different frequencies mimicking the MPR measurements. The model has only two unknown parameters $k_{eff}$ and $D_{air}$ to be fitted. We notice that $|\theta_s|$ has different sensitivities to $k_{eff}$ and $D_{air}$ at different frequencies, which is the basis for obtaining them simultaneously with a frequency sweep from 0.03 to 20 Hz. $|\theta_s|$ is more sensitive to $k_{eff}$ at low frequency and to $D_{air}$ at intermediate frequency. By fitting the model to the data using the Levenberg-Marquardt method [34] [Fig. 1(d) and Supplemental Material [29], S6], both $k_{eff}$ and $D_{air}$ of the granular flows were determined.

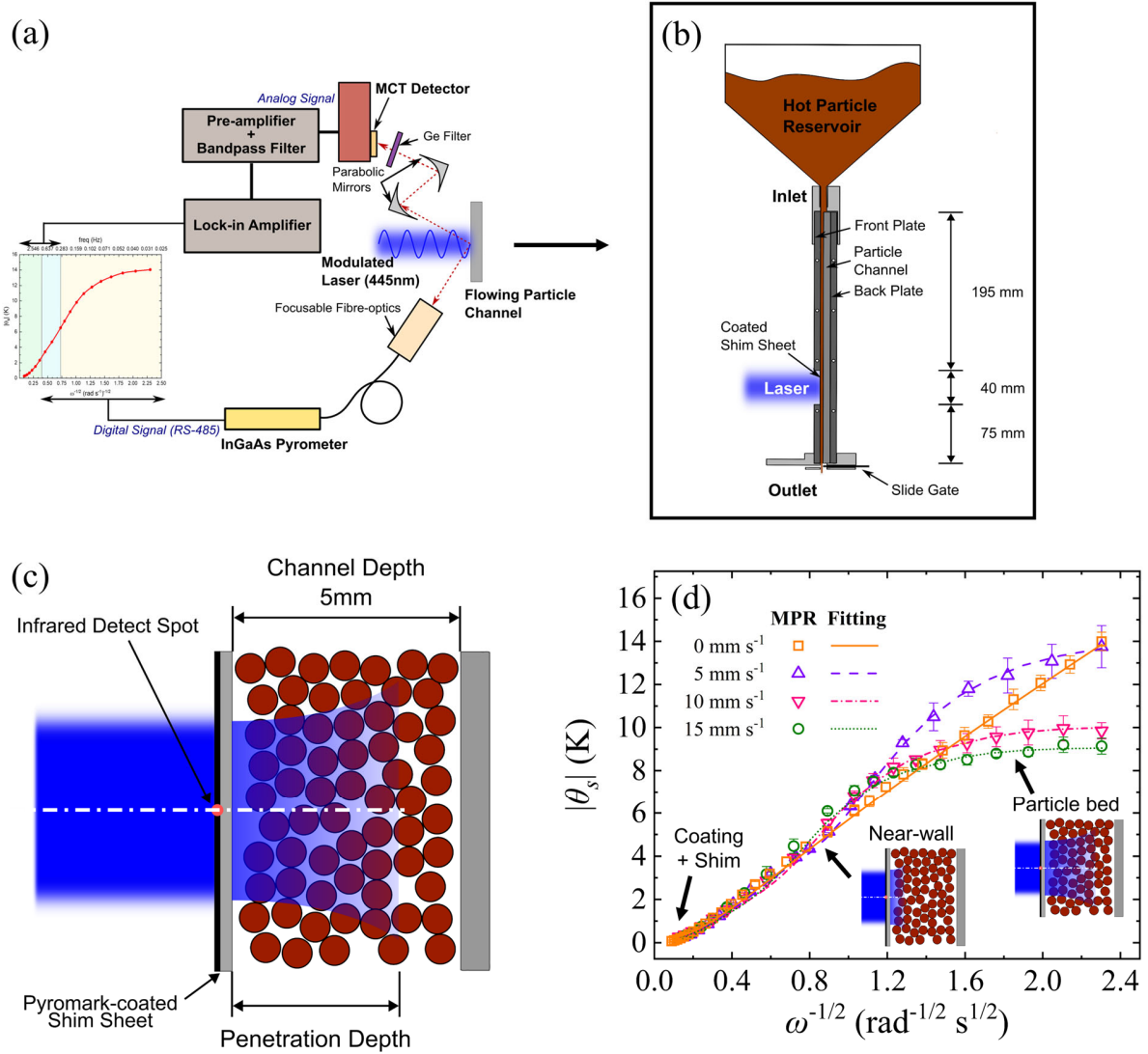

FIG. 1. Overview of MPR measurement on a flowing particle bed. (a) MPR signal collecting system. (b) Flowing particle bed in a vertical channel. (c) Schematic of laser heat penetration into the particle bed. (d) HSP 40/70 experimental data and fitting at 300 ºC; error bars are standard deviations of replicate measurements. Insets: thermal penetration depth at different frequencies.

Figs. 2(a) and 2(b) show the best-fitted $k_{eff}$ of stationary and flowing CP 40/100 and HSP 40/70, respectively. From 0 mm s$^{-1}$ to 15 mm s$^{-1}$, there is a reduction of 14% - 26% in $k_{eff}$ for both types of particles. This phenomenon has not been accounted for in previous works [1,16,17] using $k_{eff}$ of stationary beds to model flowing particles, presumably because $k_{eff}$ of flowing particle beds could not be separated from the $R_{NW}$ in traditional measurements. Besides, the $k_{eff}$ in granular flows shows a notable dependency on temperature. From 300 ºC to 650 ºC, $k_{eff}$ of CP 40/100 and HSP 40/70 beds increase by around 39% and 21%, respectively. This is mainly due to the enhanced gaseous conduction at elevated temperature, which plays an important role in particle-particle heat transfer [27]. A stronger radiative conduction also contributes to the higher $k_{eff}$.

Figs. 2(c) and 2(d) show the extracted $D_{air}$ from MPR experiments. At the stationary state, CP 40/100 and HSP 40/70 have $D_{air}$ of 14.6 μm and 20.5 μm respectively, or about 5% of their respective particle diameters. This non-zero air gap in the stationary beds has not been revealed in prior measurements because of the insensitivity of the techniques in the near-wall region. The air gap can be attributed to an average particle-wall distance since the local packing density immediately near the wall is always zero for spheres [19,35]. When particles started flowing, these air gap thicknesses increased to approximately 31 μm for both types of particles. The $D_{air}$ in the flowing granular media is about $(0.08 - 0.11)D_p$, which is comparable to the literature ($D_{air} \sim D_p/10$) [1,14,16,36]. Unlike $k_{eff}$, the $D_{air}$ values show little temperature dependence from 300 ºC to 650 ºC. This is because $D_{air}$ is mainly determined by mechanical properties of granular flows and the wall with weak temperature dependence [37].

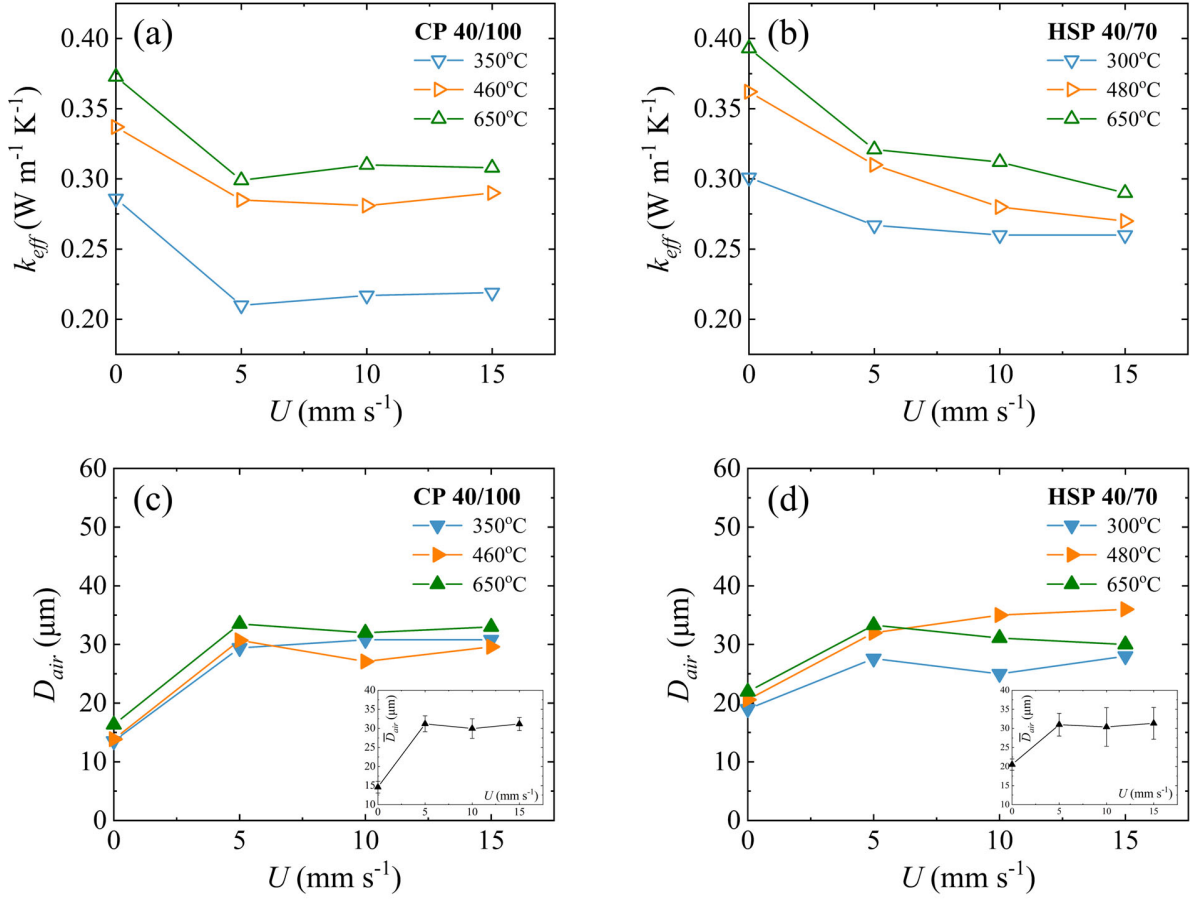

FIG. 2. Model fitting results of flowing beds as a function of velocity. $k_{eff}$ of (a) CP 40/100 and (b) HSP 40/70. $D_{air}$ of (c) CP 40/100 and (d) HSP 40/70. Insets show the averaged $D_{air}$ over three temperatures with error bars representing the standard deviation.

Since we have not observed any significant difference in the average packing density between stationary and flowing particle beds (see Supplemental Material [29], S2), the origins of the $k_{eff}$ reduction and $D_{air}$ increase can be related to the particle packing structure in the bulk and near-wall regions respectively. To reveal these, we employed DEM to simulate flowing particle beds by using LIGGGHTS [LAMMPS (Large-scale Atomic/Molecular Massively Parallel Simulator) Improved for General Granular and Granular Heat Transfer Simulations] [38]. Even though the Carbo particles are spheroids, they were assumed to be spherical in DEM simulations. Earlier studies did show that the particle shape can impact thermal conductivity of stationary particle beds [39] and granular flows [40,41] through variations in contact area ratio, interstitial spaces for gas conduction and radiation, and packing structures in the near-wall and bulk regions. In our case, however, the particles having roundness of 0.8 means the behavior is expected to be similar to that of spherical particles. Furthermore, the shape effect was found to be weak for smaller diameters such as ~ 400 μm studied here [39]. As shown in Fig. 3(a), a particle reservoir and a rectangular channel similar to our experimental setup were defined as the simulation domain. The total number of particles was over 45,000, large enough to simulate the experiments. The cross-sectional area of the channel was chosen to be $15D_p \times 15D_p$, also similar to the experimental setup. Particles generated in the reservoir will first pile up from bottom and fill the entire chamber due to gravity. Then the particle flow can be stabilized at a controllable velocity by regulating the opening size of the bottom outlet. Both CP 40/100 and HSP 40/70 were simulated and all mechanical properties were based on experimental data of CARBOBEAD CP 30/60 with the same composition and similar particle diameter of 426 μm [37,42]. The mechanical properties have weak temperature dependence within the temperature range considered here [42] and were assumed to be constant in the simulation. The properties and parameters used in DEM simulations are listed in Table 1. The Hertz model was implemented for modeling particle-particle and particle-wall interaction at contact points [43,44], and an alternative elastic-plastic spring-dashpot model for rolling friction was applied due to its universality in most of particle settling problems [44,45].

TABLE I. DEM simulation parameters.

| Parameter (unit) | Value |
| --- | --- |
| $D_p$ ($\mu m$) | 275, 400 |
| Skin Distance ($\mu m$) | $D_p/2$ |
| Time Step (sec) | $8 \times 10^{-9}$ |
| Young's Modulus (GPa) | 240 |
| Poisson's Ratio | 0.26 |
| Coefficient of Restitution | 0.5 |
| Coefficient of Friction | 0.59 |
| Coefficient of Rolling Friction | 0.28 |
| Particle Density ($kg\ m^{-3}$) | 3480 |

To obtain $k_{eff}$, the real packing structures in particle beds were extracted from DEM simulations and fed into COMSOL models as shown in Fig. 3(a). A $10D_p \times 8D_p \times 1.3D_p$ slab adjacent to the channel wall and a $5D_p \times 5D_p \times 5D_p$ cube from the center of channel were selected to represent the near-wall and bulk regions respectively. Both domains are sufficiently large to average out spatial difference in packing structure [27]. The $k_{eff}$ of the cube in Fig. 3(b) can be calculated in COMSOL by integrating the heat flux $q$ over plane $S$ after applying a temperature gradient $\Delta T/L$ across the cube:

$$k_{eff} = \frac{L}{\Delta T}\frac{\iint q\, dS}{S} \qquad (2)$$

The $k_{eff}$ of stationary ceramic particle beds at high temperature has been measured using a transient hot-wire (THW) method in the previous study [3]. As shown in Fig. 3(c), the THW value of $k_{eff}$ for stationary HSP 40/70 bed at 300 ºC is 27% higher than our present MPR result at zero velocity. This deviation comes from the non-uniform packing density distribution across the particle bed [Fig. 3(d)]. The THW method measures the center of a packed particle bed where the packing is denser with a nearly constant solid fraction of 58%. In contrast, the MPR measures the region about 1 mm (or about $3D_p$) from the wall based on the thermal penetration depth [Eq. (1)] in the experiments, where the packing density experiences a large fluctuation and within $1.5D_p$, a sharp decrease to zero towards the wall [Fig. 3(d)]. The different measurement locations between MRP and THW lead to different stationary $k_{eff}$ values, as also well captured by our DEM+COMSOL simulations [Fig. 3(c)].

With an increasing flow velocity, results from DEM+COMSOL simulations exhibit a similar decreasing trend of near-wall $k_{eff}$ in the granular flow, consistent with values measured by the MPR [Fig. 3(c)]. The DEM simulation was validated by accurately modeling $k_{eff}$ as a function of flow velocity and location. We then used it to extract the parameters characterizing the packing structure and particle behavior in granular flow. We defined the *coordination number* as the number of surrounding particles in contact with the central particle [Fig. 3(e)]. When the flow velocity in DEM simulations increases from 0 to 35 mm s$^{-1}$, the average coordination number decreases monotonically from 1.8 to 1.1, indicating less contact points in flowing particles. This structural change indicates the dilation induced by shear [46,47]. Although this dilation does not cause noticeable variation in the packing density, the reduction of the number of particle-particle heat conduction pathways will result in a notably decreasing $k_{eff}$.

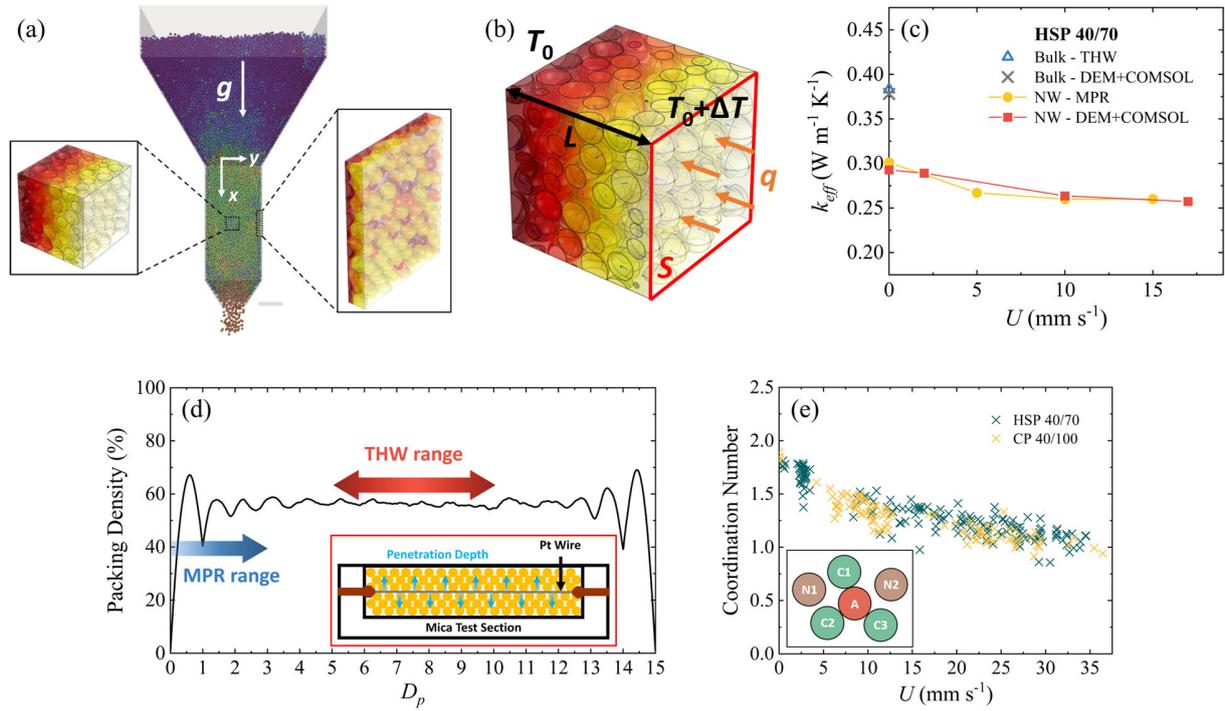

FIG. 3. DEM simulation results of packing structure's impact on $k_{eff}$. (a) A snapshot of DEM simulation and particle bed samples extracted from the bulk and the near-wall region. (b) Modeling of $k_{eff}$ in COMSOL. (c) Comparison between MPR results, THW results and simulated $k_{eff}$ at different locations in the HSP 40/70 bed at 300 ºC. (d) Wall-to-wall packing density distribution from DEM simulation. Inset: a schematic of THW measurement. (e) Average coordination number of granular media. The inset shows particle A having a coordination number of 3; letters $C$ and $N$ represent particles in contact and not in contact with particle A, respectively.

Moreover, we seek to understand the micromechanical origin of the over 50% increase in $D_{air}$ observed in flowing particles compared to the stationary case. In dense granular flows with frequent semi-inelastic collisions and frictions, the particle motion is diffusive at the time scale similar to $D_p/U$ (a particle travelling the distance of $D_p$) [20,48-51]. It is reasonable to attribute the origin of the air gap to particle diffusion transverse to the flow direction in the near-wall region where particles have higher mobility and may more easily diffuse away from the wall. As a measure of the mobility, particle velocity fluctuation can be calculated by:

$$v_i' = \sqrt{\frac{\sum_{k=1}^{N}(v_{i,k} - \bar{v}_i)^2}{N}} \tag{3}$$

where $v_{i,k}$ is the velocity component in the $i$ direction of the $k^{\text{th}}$ particle, and $\bar{v}_i$ is the average of $v_{i,k}$ among $N$ particles studied in DEM simulations. Since the heat transfer and particle behavior of interest is in the transverse direction ($y$ direction) [Fig. 3(a)], only properties in this direction were analyzed. Based on the dense-gas kinetic theory [52], the transverse self-diffusivity $D_{yy}$ is calculated by:

$$D_{yy} = \frac{D_p \left(\frac{\pi v_y'}{3}\right)^{\frac{1}{2}}}{8(1+e_p)\nu g_0(\nu)} \tag{4}$$

where $e_p$ is the coefficient of restitution of particles, $\nu$ is the solid fraction, and $g_0(\nu)$ is an equation of state of ridge spheres given by the Carnahan-Starling expression [53]:

$$g_0(\nu) = \frac{2-\nu}{2(1-\nu)^3} \tag{5}$$

As seen in Figs. 4(a) and 4(b), the $D_{yy}$ peaks near the wall and decreases towards the center as the wall gradually transmits shear work into the particle bed [22]. Higher flow velocity will induce stronger particle-wall interaction and a larger local shear rate, resulting in increasing $D_{yy}$ of the first layer of particles adjacent to the wall [Fig. 4(c)]. This shear-rate dependent $D_{yy}$ has also been observed both experimentally [22,51] and numerically [54]. During the time span of $D_p/U$, the transverse mean square displacement in the first layer of wall-adjacent particles can be estimated as its diffusion length scale $L_D$ by [50]:

$$L_D = \sqrt{D_{yy}|_{y=0.5D_p} \frac{D_p}{U}} \tag{6}$$

We compared $L_D$ and the measured increment in $D_{air}$ between stationary and flowing particle beds. As shown in Fig. 4(d), the $D_{air}$ predicted from the simple scaling equation of Eq. (6) and that extracted from the MPR measurements [Figs. 2(c) and 2(d)] have excellent agreement. We can conclude that the dense-gas kinetic theory combined with DEM simulations can well capture the $R_{NW}$. This finding thus reveals that the increasing $D_{air}$ in flowing particle beds is a result of the enhanced diffusive motion of particles in the near-wall region.

One minor fact to note is that $v_y'$ peaks at $y = D_p$ but not $y = 0.5D_p$, which was also observed in experiments by Natarajan and Hunt [18,22]. In granular flows, $y = 0.5D_p$ and $y = D_p$ correspond to the first and second layers of particles adjacent to the wall, respectively. Although the first layer is impacted to the maximum extent by interactions with the wall, it has a lower $v'_y$ because a non-negligible fraction of shear work is converted into rotational kinetic energy ($E_{k,r}$). During the collisions between the first and second layers, the $E_{k,r}$ is converted back to translational kinetic energy ($E_{k,t}$), leading to the maximum $v'_y$ at $y = D_p$. This is clearly shown in the plots of average $E_{k,r}$ and $E_{k,t}$ obtained from DEM simulations (see Supplemental Material [29], S7).

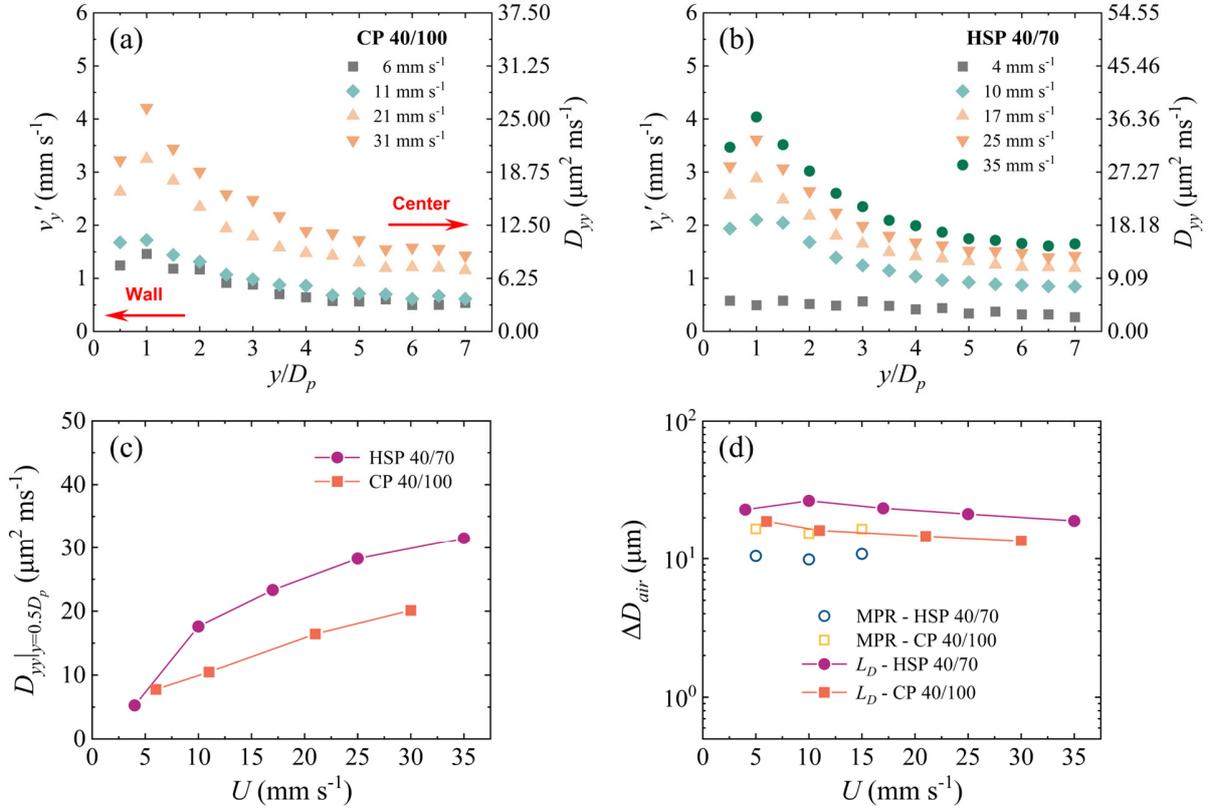

FIG. 4. DEM simulation results of particle diffusive behavior. $v'_y$ and $D_{yy}$ of flowing (a) CP 40/100 and (b) HSP 40/70. (c) $D_{yy}$ of the first layer of wall-adjacent particles. (d) Comparison of the $L_D$ and the measured increment in $D_{air}$ between stationary and flowing particle beds.

We further examined the following probability $P_{center}$ to show the directional preference of particle motion (Fig. 5):

$$P_{center} = \frac{N_c}{N_c + N_w} \tag{7}$$

where $N_c$ and $N_w$ are the numbers of particles with their $v_{y,k}$ towards the channel center and the wall, respectively. Interestingly, DEM simulations show that particles in the near-wall region are more likely to move towards the center ($P_{center} > 0.5$). Campbell [55] also observed that particles may move in preferred directions due to certain packing structures induced by shear motion. These results all imply a larger particle-wall separation in flowing particles, which in our measurement is manifested as increasing air gaps when particles are flowing.

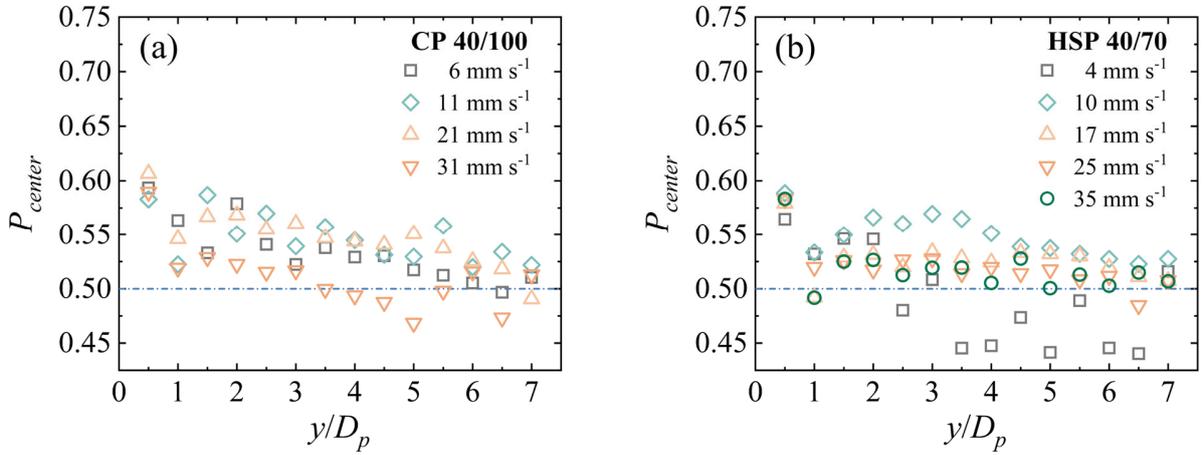

FIG. 5. $P_{center}$ of flowing (a) CP 40/100 and (b) HSP 40/70.

In summary, via a unique non-contact frequency-domain measuring technique (MPR) to probe into gravity-driven granular flows in a vertical channel, we are able to separately quantify the near-wall resistance and the bulk thermal conductivity of particle beds. We observed that as the particles are flowing, the near-wall air gap ($D_{air}$) and resistance increase while the bulk effective thermal conductivity ($k_{eff}$) decreases, both of which point to weakened heat transfer from the wall to the granular flow. Combined with DEM simulations, we examined the microscale changes in packing feature of granular flows. The increasing particle dilation and particle-wall separation at higher flow velocity accounts for the decreasing $k_{eff}$ and increasing $D_{air}$, respectively. Instead of elaborating the particle-wall separation via Natarajan and Hunt's analytical model based on several assumptions [18], we well predicted the $D_{air}$ increase using the dense-gas kinetic theory with self-diffusivities obtained from DEM simulations, which reveals the importance of the wall shear causing particles in the wall-adjacent layer moving away from the wall. This work provides fundamental understanding of the microscopic picture of heat transfer across granular flows. We believe the presented experimental and simulation results can motivate more comprehensive model to precisely depict the behavior of flowing particles.

**Acknowledgements**

This material is based upon work supported by the U.S. Department of Energy's Office of Energy Efficiency and Renewable Energy (EERE) under Solar Energy Technologies Office (SETO) Agreement Number DE-EE0008379. The views expressed herein do not necessarily represent the views of the U.S. Department of Energy or the United States Government.

**References**

[1]  K. J. Albrecht and C. K. Ho, Journal of Solar Energy Engineering **141**, 031006 (2019).
[2]  J. D. D. Hertel and S. Zunft, Appl Therm Eng **206** (2022).
[3]  K. M. Chung, J. Zeng, S. R. Adapa, T. Feng, M. V. Bagepalli, P. G. Loutzenhiser, K. J. Albrecht, C. K. Ho, and R. Chen, Solar Energy Materials and Solar Cells **230**, 111271 (2021).
[4]  C. H. Rycroft, G. S. Grest, J. W. Landry, and M. Z. Bazant, Physical review E **74**, 021306 (2006).
[5]  B. Yohannes, H. Emady, K. Anderson, I. Paredes, M. Javed, W. Borghard, F. Muzzio, B. Glasser, and A. Cuitiño, Physical Review E **94**, 042902 (2016).
[6]  F. L. Qi and M. M. Wright, Powder Technol **335**, 18 (2018).
[7]  R. W. Houim and E. S. Oran, Journal of fluid mechanics **789**, 166 (2016).
[8]  Q. Yang, A. S. Berrouk, Y. Du, H. Zhao, C. Yang, M. A. Rakib, A. Mohamed, and A. Taher, Applied Mathematical Modelling **40**, 9378 (2016).
[9]  S. Savage, Journal of Fluid Mechanics **377**, 1 (1998).
[10] F. Chevoir, M. Prochnow, P. Moucheront, F. da Cruz, F. Bertrand, J.-P. Guilbaud, P. Coussot, and J.-N. Roux, in *Powder and Grains 2001* (CRC Press, 2020), pp. 399.
[11] GDRMiDi, The European Physical Journal E **14**, 341 (2004).
[12] S. Hsiau and M. Hunt, Journal of Fluid Mechanics **251**, 299 (1993).
[13] P. Rognon and I. Einav, Physical review letters **105**, 218301 (2010).
[14] W. N. Sullivan and R. Sabersky, International Journal of Heat and Mass Transfer **18**, 97 (1975).
[15] A. Denloye and J. Botterill, Chemical Engineering Science **32**, 461 (1977).
[16] Y. Yu, F. Nie, F. Bai, and Z. Wang, International Journal of Heat and Mass Transfer **180**, 121725 (2021).
[17] G. Wei, P. Huang, L. Pan, L. Cui, C. Xu, and X. Du, International Journal of Heat and Mass Transfer **187**, 122571 (2022).
[18] V. Natarajan and M. Hunt, International journal of heat and mass transfer **41**, 1929 (1998).

[19] V. Natarajan and M. Hunt, EXPERIMENTAL HEAT TRANSFER An International Journal **10**, 89 (1997).
[20] N. Menon and D. J. Durian, Science **275**, 1920 (1997).
[21] S. Moka and P. R. Nott, Physical review letters **95**, 068003 (2005).
[22] V. Natarajan, M. Hunt, and E. Taylor, Journal of Fluid Mechanics **304**, 1 (1995).
[23] M. F. Watkins, Y. N. Chilamkurti, and R. D. Gould, Journal of Heat Transfer **142**, 022103 (2020).
[24] T. A. Rulko, B. J. Li, B. Surhigh, J. M. Mayer, and R. B. Chandran, Sol Energy **264** (2023).
[25] J. Zeng, K. M. Chung, Q. Wang, X. Wang, Y. Pei, P. Li, and R. Chen, International Journal of Heat and Mass Transfer **170**, 120989 (2021).
[26] J. Zeng, K. M. Chung, S. R. Adapa, T. Feng, and R. Chen, International Journal of Heat and Mass Transfer **180**, 121767 (2021).
[27] J. Zeng, K. M. Chung, X. Zhang, S. Adapa, T. Feng, Y. Pei, and R. Chen, Journal of Applied Physics **130** (2021).
[28] J. Zeng, K. M. Chung, X. Zhang, T. Feng, S. Adapa, and R. Chen, Annual Review of Heat Transfer **25** (2022).
[29] See Supplemental Material for additional details on experiments and models, which includes Refs. [3,25,30-34].
[30] C. S. Kim, Thermophysical properties of stainless steels, 1975.
[31] A. C. Yunus, *Fluid Mechanics: Fundamentals And Applications (Si Units)* (Tata McGraw Hill Education Private Limited, 2010).
[32] N. P. Siegel, M. D. Gross, and R. Coury, Journal of Solar Energy Engineering **137**, 041003 (2015).
[33] https://www.mcmaster.com/products/sheets/shim-stock-6/.
[34] B. Cheng, B. Lane, J. Whiting, and K. Chou, Journal of manufacturing science and engineering **140**, 111008 (2018).
[35] E. Schlunder, in *International Heat Transfer Conference Digital Library* (Begel House Inc., 1982).
[36] J. Spelt, C. Brennen, and R. Sabersky, International Journal of Heat and Mass Transfer **25**, 791 (1982).
[37] M. V. Bagepalli, J. D. Yarrington, A. J. Schrader, Z. M. Zhang, D. Ranjan, and P. G. Loutzenhiser, Solar Energy **207**, 77 (2020).
[38] C. Kloss, C. Goniva, A. Hager, S. Amberger, and S. Pirker, Progress in Computational Fluid Dynamics, an International Journal **12**, 140 (2012).
[39] J. Q. Gan, Z. Y. Zhou, and A. B. Yu, Powder Technol **311**, 157 (2017).
[40] B. Nadler, F. Guillard, and I. Einav, Phys Rev Lett **120** (2018).
[41] P. W. Cleary, Powder Technol **179**, 144 (2008).
[42] J. D. Yarrington, M. V. Bagepalli, G. Pathikonda, A. J. Schrader, Z. M. Zhang, D. Ranjan, and P. G. Loutzenhiser, Sol Energy **213**, 350 (2021).
[43] A. Di Renzo and F. P. Di Maio, Chemical engineering science **60**, 1303 (2005).
[44] J. Ai, J.-F. Chen, J. M. Rotter, and J. Y. Ooi, Powder Technology **206**, 269 (2011).
[45] K. Iwashita and M. Oda, Journal of engineering mechanics **124**, 285 (1998).
[46] K. Krishnaraj and P. R. Nott, Nature communications **7**, 10630 (2016).
[47] K. Sakaie, D. Fenistein, T. J. Carroll, M. van Hecke, and P. Umbanhowar, Europhysics Letters **84**, 38001 (2008).

[48] S.-S. Hsiau and Y.-M. Shieh, Journal of Rheology **43**, 1049 (1999).
[49] A. Sierou and J. F. Brady, Journal of fluid mechanics **506**, 285 (2004).
[50] J. Choi, A. Kudrolli, R. R. Rosales, and M. Z. Bazant, Physical review letters **92**, 174301 (2004).
[51] B. Utter and R. P. Behringer, Physical Review E **69**, 031308 (2004).
[52] S. Savage and R. Dai, Mechanics of Materials **16**, 225 (1993).
[53] N. F. Carnahan and K. E. Starling, The Journal of chemical physics **51**, 635 (1969).
[54] Y. Fan, P. B. Umbanhowar, J. M. Ottino, and R. M. Lueptow, Physical review letters **115**, 088001 (2015).
[55] C. S. Campbell, Journal of Fluid Mechanics **348**, 85 (1997).

**Supplemental Material for "Micromechanical Origin of Heat Transfer to Granular Flow"**

Xintong Zhang[*,1], Sarath Adapa[*,1], Tianshi Feng[1], Jian Zeng[1], Ka Man Chung[2], Clifford Ho[3], Kevin Albrecht[3], Renkun Chen[#1,2]

[1]Department of Mechanical and Aerospace Engineering, University of California, San Diego, La Jolla, California 92093, United States

[2]Program in Materials Science and Engineering, University of California, San Diego, La Jolla, California 92093, United States

[3]Concentrating Solar Technologies Department, Sandia National Laboratories, 1515 Eubank Blvd. SE, Albuquerque, New Mexico, 87123, United States

[*]These authors contributed equally to this work

[#]Corresponding Author: rkchen@ucsd.edu

# S1. Photographs of MPR system for granular flow measurement.

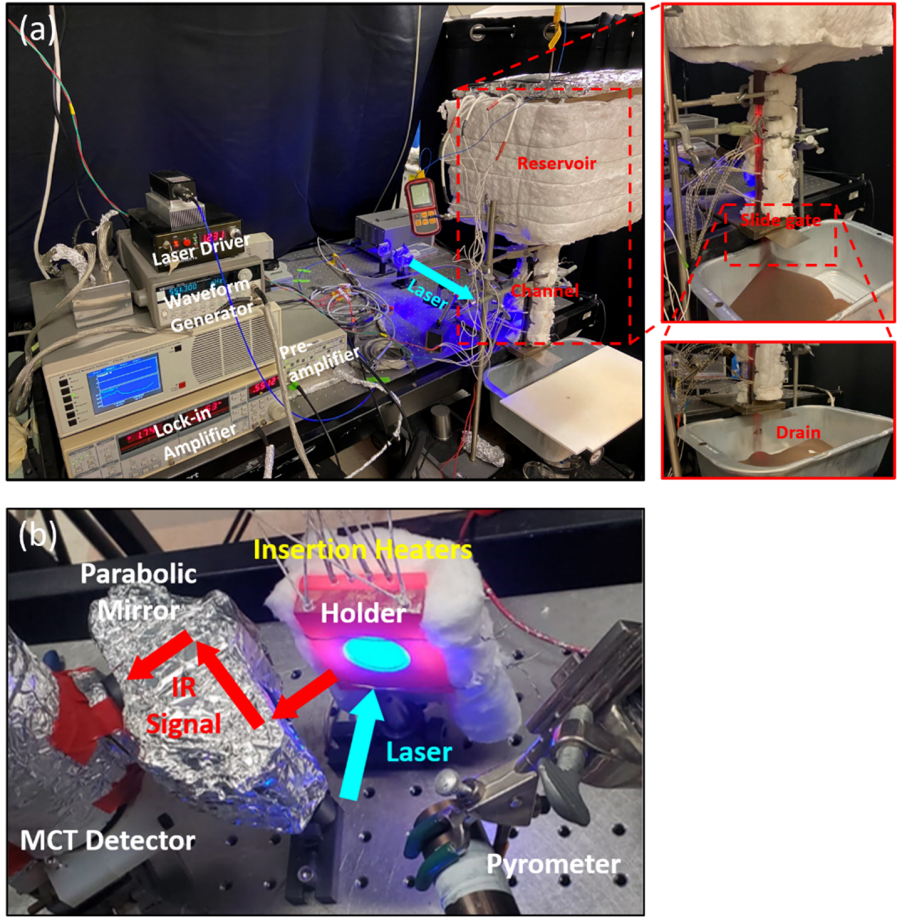

FIG. S1. Photos of the MPR measurement system for (a) granular flows confined in a 5 mm-deep vertical channel with a heated particle reservoir on the top and a slide gate at the bottom and (b) stationary particle bed confined in a particle holder with a cavity. Both of the channel and the holder are red when heated up to 650 °C.

# S2. Particle bed effective density measurement

As shown in Fig. S2(a), the front cover of the channel is replaced by a transparent acrylic cover to observe the particle level and measure the effective density of particle bed $\rho_{bed}$. The channel width is 30 mm and the channel depth is adjustable from 2 mm to 7.5 mm. The height of particle level is measured via the scale on the acrylic cover. The flow velocity $U$ is calculated by

$$U = \frac{\Delta h}{t} \qquad (S1)$$

where $\Delta h$ is the displacement of particle level within the time interval $t$. With the knowledge of the channel cross-sectional area $A$, the volumetric flow rate $\dot{V}$ is calculated by

$$\dot{V} = UA \qquad (S2)$$

and the mass flow rate $\dot{m}$ is calculated by

$$\dot{m} = \frac{m}{t} \qquad (S3)$$

where $m$ is the mass of collected particles on the balance within the time interval $t$. Then $\rho_{bed}$ can be calculated by

$$\rho_{bed} = \frac{\dot{m}}{\dot{V}} \qquad (S4)$$

We measured $\rho_{bed}$ of HSP 40/70 with different channel depths. Fig. S2(b) shows that $\rho_{bed}$ does not depend on $U$ when $U$ is below 35 mm s$^{-1}$.

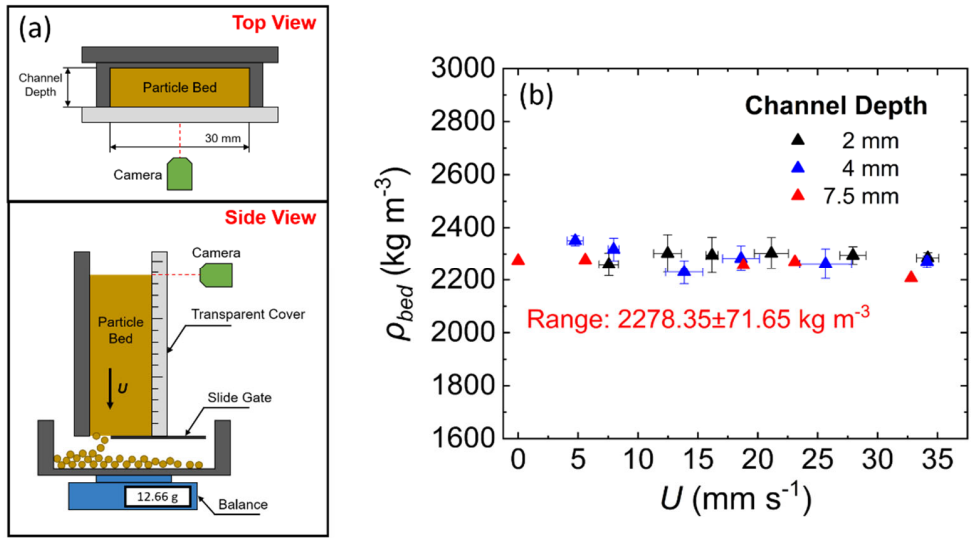

FIG. S2. Measurement of effective density of particle bed. (a) Schematics of the measurement system with the same configuration as the MPR setup and a transparent acrylic cover. (b) $\rho_{bed}$ of HSP 40/70 as a function of $U$ with different channel depths.

## S3. Calibration of Laser Heat Flux

Borosilicate glass was chosen as the reference material for calibration for its well characterized thermophysical properties (effusivity $e = \sqrt{\rho c k}$ for MPR measurements, where $\rho$, $c$, and $k$ are the density, the specific heat, and the thermal conductivity of sample, respectively) and stability at high temperatures. The same process used to coat the shim sheet was also used to coat the reference sample with Pyromark 2500 paint to ensure similar thicknesses of the laser absorption layer. Since the measurements were sensitive to laser heat flux, the optics were checked for alignment each time and the laser flux was calibrated every two weeks. Fig. S3(a) shows the voltage signal $V_{rms}$ obtained from the MCT detector during one such calibration procedure on the reference Borosilicate sample. A factory calibrated pyrometer was used to simultaneously record the amplitude of surface temperature oscillation $|\theta_s|$ in the frequency range of 0.5 – 3 Hz. As shown in Fig. S3(b), this $|\theta_s|$ was used to reference $V_{rms}$ based on a linear relationship. In this way, we can use the MCT detector to obtain $|\theta_s|$ at higher frequencies out-of-range for the pyrometer. Subsequently, we can obtain the heat flux from the slope of the $|\theta_s|$ vs. $1/\sqrt{\omega}$ curve (the slope equals to $\frac{1}{e}$ numerically) and comparing it to literature values of the borosilicate sample. More details can be found in Ref. [1].

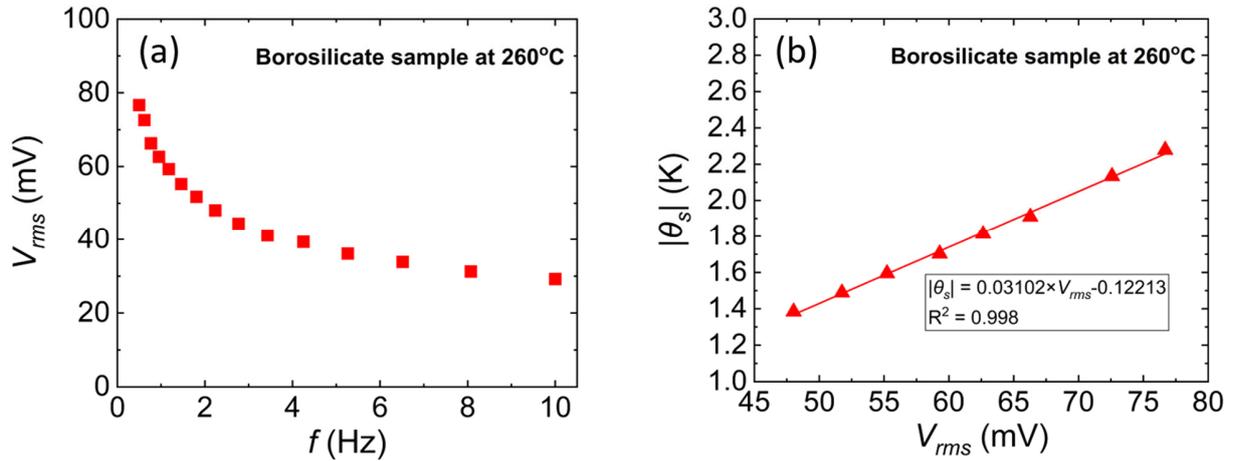

FIG. S3. (a) Voltage signal from the MCT detector during calibration procedure on the borosilicate reference sample. (b) Linear relationship used to reference MCT voltage signal to the amplitude of surface temperature oscillations measured from the pyrometer.

**S4. Surface morphology and size distribution of CARBO ceramic particles**

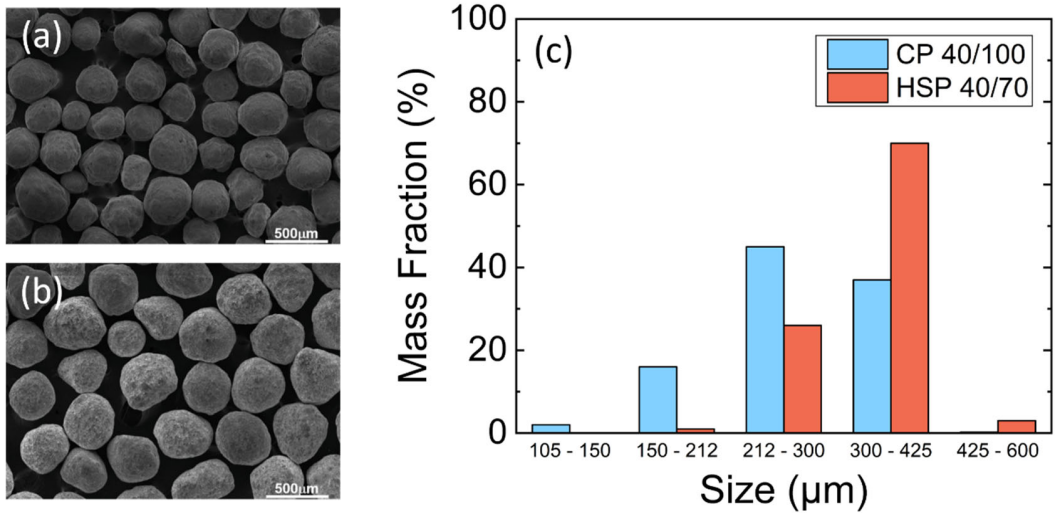

FIG. S4. Scanning electron microscopy (SEM) images of (a) CP 40/100 (mean diameter: 275 μm) and (b) HSP 40/70 (mean diameter: 404 μm) [2]. (c) Size distribution of CP 40/100 and HSP 40/70.

**S5. Proof of plug flow assumption of granular flow via DEM simulation**

Fig. S5(a) and S5(b) show that flowing CP 40/100 and HSP 40/70 particle beds below 17 mm s$^{-1}$ are plug flow with a relatively uniform velocity profile via DEM. Velocity profiles of flowing particle beds at higher velocities have a deviation from plug flow at channel wall but are still close to it in the bulk region. This demonstrates the validity of the plug flow assumption when modeling the flowing particle bed at velocity lower than 15 mm s$^{-1}$ in our MPR measurements.

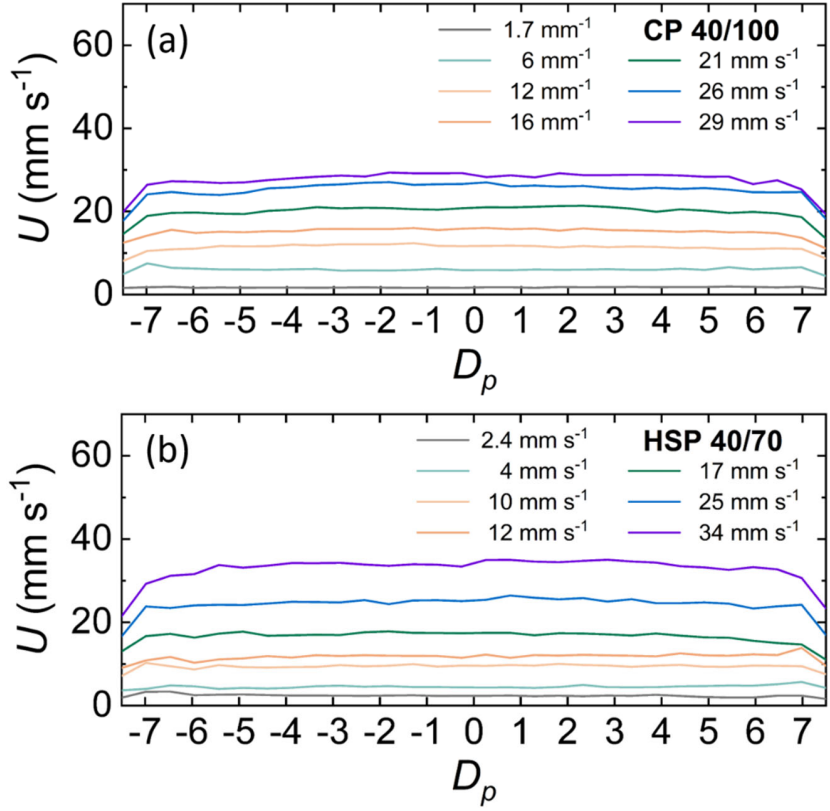

FIG. S5. DEM simulated velocity distribution across the channel for (a) CP 40/100 and (b) HSP 40/70.

**S6. Details on COMSOL Multiphysics® model fitting**

**(I) Introduction of the COMSOL Multiphysics® model**

A coupled conduction and convection model was built in COMSOL Multiphysics® using the Heat Transfer in Solids and Fluids module. The governing equations in different domains are given by

In fluid:

$$\rho_{bed} c_{bed} \frac{\partial T}{\partial t} + \rho_{bed} c_{bed} U \nabla T + \nabla(-k_{eff} \nabla T) = q_0 \qquad (S5)$$

In stagnant materials:

$$\rho_s c_s \frac{\partial T}{\partial t} + \nabla(-k_s \nabla T) = q_0 \qquad (S6)$$

where $\rho$, $c$, and $k$ are density, specific heat and thermal conductivity of material; subscripts $bed$ and $s$ represents the particle bed and stagnant materials. As shown in Fig. S6, we set the flowing particle bed as homogeneous continuum with constant thermophysical properties and a plug flow profile. The incident heat flux $q_0$ from the laser calibration was applied on the front surface with an angular frequency $\omega$. To account for the radiation heat loss from the black paint coating, a surface-to-ambient radiation heat loss term was added with a coating emissivity of 0.9. An effective air gap with thickness $D_{air}$ was defined as a stagnant layer between the flowing particles and the inner surface of alloy sheet. A time-dependent solver was used to simulate the MPR experiments.

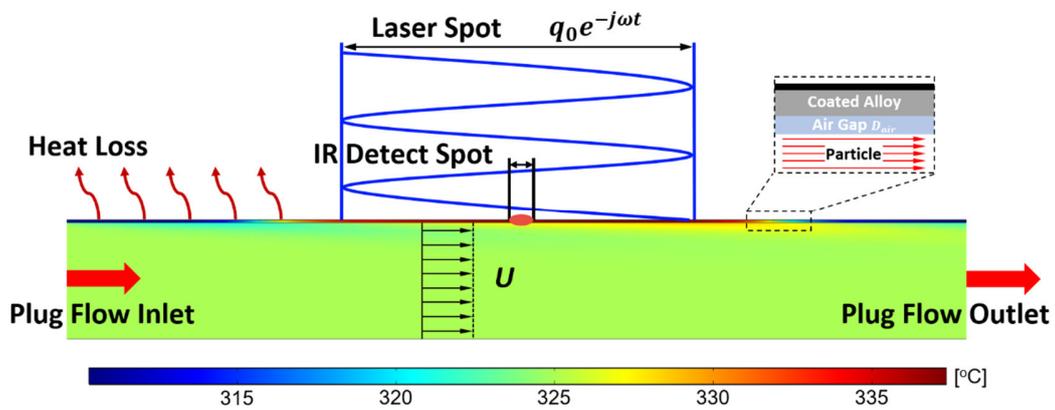

FIG. S6. Schematic of modeled MPR measurements on flowing particle beds in COMSOL Multiphysics®.

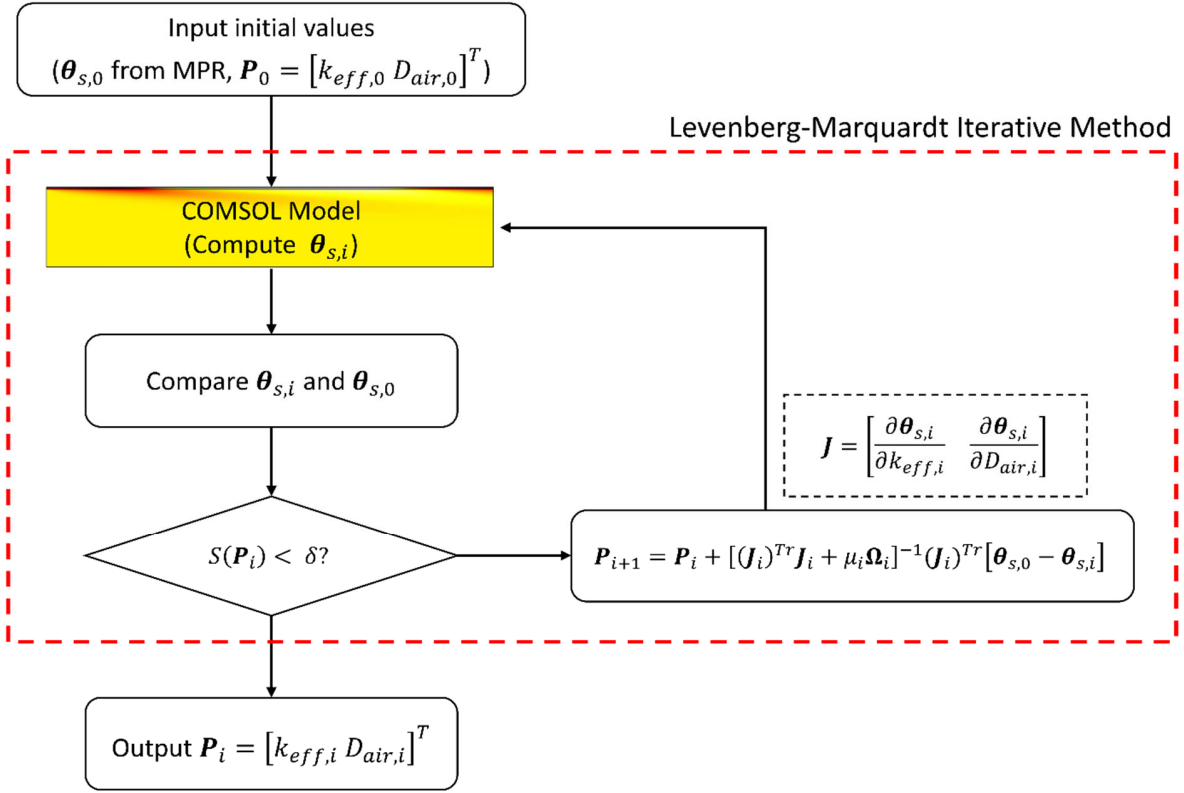

FIG. S7. Levenberg-Marquardt iterative method for simultaneously fitting both $k_{eff}$ and $D_{air}$ via the COMSOL Multiphysics® model.

Levenberg-Marquardt (L-M) iterative method is a standard method for the fitting of inverse heat transfer problems [3]. As shown in Fig. S7, our COMSOL Multiphysics® model was combined with the (L-M) algorithm to extract the values of $k_{eff}$ and $D_{air}$. Here, vector $\boldsymbol{P}$ contains the fitting parameters $k_{eff}$ and $D_{air}$. Vector $\boldsymbol{\theta}_s$ contains the values of $|\theta_s|$ in the full frequency range. The superscript $i$ represents the $i^{th}$ iteration step. At each iteration step, $\boldsymbol{\theta}_{s,i}$ is calculated by the COMSOL Multiphysics® model. The error between the simulated $\boldsymbol{\theta}_{s,i}$ and measured $\boldsymbol{\theta}_{s,0}$ from MPR is estimated by

$$S(\boldsymbol{P}_i) = |\boldsymbol{\theta}_{s,i} - \boldsymbol{\theta}_{s,0}|^2 \tag{S7}$$

The convergence criterion was established by comparing the errors of two consecutive iteration steps. If the subsequent step has an error larger than the current step, the iteration process will be terminated and the fitting results are stored in $\boldsymbol{P}_i$. If the error in the subsequent step is lower than the current step, $\boldsymbol{P}_{i+1}$ for the next step will be obtained by

$$P_{i+1} = P_i + [(J_i)^{Tr}J_i + \mu_i \Omega_i]^{-1}(J_i)^{Tr}[\theta_{s,0} - \theta_{s-i}(P^i)] \quad (S8)$$

where $\mu_i$ is the damping factor, $\Omega_i = diag[(J_i)^{Tr}J_i]$, and the Jacobian matrix for each step ($J_i$) is numerically obtained by changing 1% of each fitting parameter in the model. Initially, a relatively large damping factor should be selected and it will gradually decrease in each iteration step. When $S(P^i)$ is lower than the allowable error limit $\delta$, the fitting process is considered as completed. Using this iterative method, both $k_{eff}$ and $D_{air}$ were obtained with a satisfying fitting quality as shown in Fig. S8.

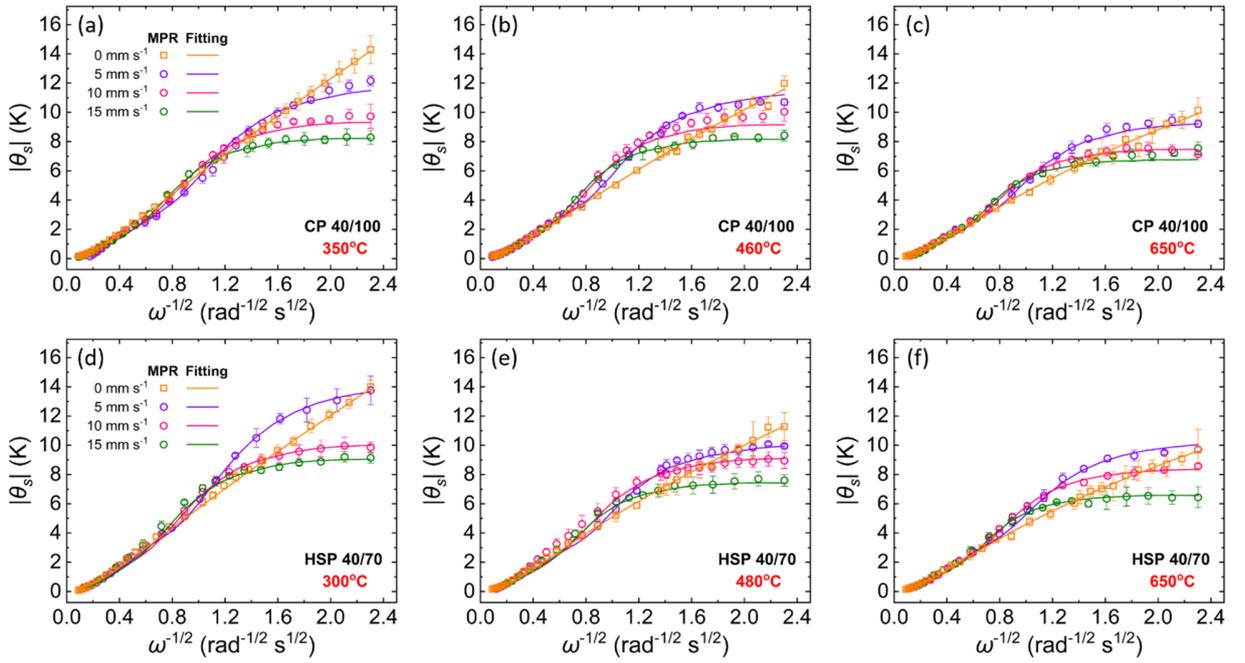

FIG. S8. Raw MPR data and corresponding COMSOL Multiphysics® model fitting curves of CP 40/100 at (a) 350 ºC, (b) 460 ºC, (c) 650 ºC and HSP 40/70 at (d) 300 ºC, (e) 480 ºC, (f) 650 ºC.

**(II) Sensitivity analysis of the COMSOL Multiphysics® model**

A sensitivity analysis has been investigated in the frequency range from 0.03 Hz to 10 Hz. The sensitivity of $|\theta_s|$ to the fitting parameter $P$ ($k_{eff}$ or $D_{air}$) is defined by

$$Sensitivity = \frac{\frac{\Delta|\theta_s|}{|\theta_s|}}{\frac{\Delta P}{P}} \quad (S9)$$

where $\Delta|\theta_s|$ is simulated by COMSOL when $\Delta P = 0.01 \times P$. Fig. S9 shows the sensitivity to $k_{eff}$ and $D_{air}$ of flowing CP 40/100 at different temperatures. Both sensitivities are close to zero at high frequency because the measured thermal response is mainly contributed by coated alloy sheet. Only when the frequency is lower than 0.1 Hz, both $k_{eff}$ and $D_{air}$ have absolute sensitivity higher than 0.2 to guarantee enough fitting resolution of separating different parameters. This sensitivity analysis also serves as a proof that the MPR technique has the capability of isolate $k_{eff}$ and $D_{air}$ in the frequency range used in our measurements.

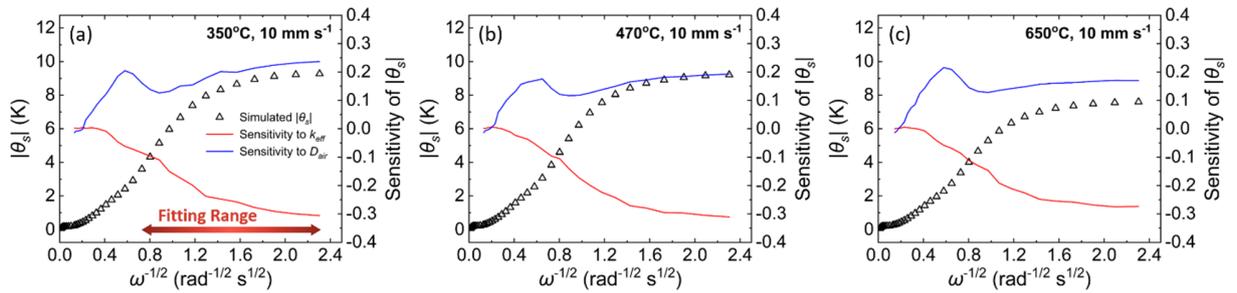

FIG. S9. Sensitivity of $|\theta_s|$ to $k_{eff}$ and $D_{air}$ in COMSOL model of flowing CP 40/100 bed with a velocity of $U$ = 10 mm s$^{-1}$. The flowing particle bed is at (a) 350 °C, (b) 470 °C, and (c) 650 °C.

### (III) Parameters in the COMOSL model

TABLE S1. Simulation Parameters in the COMSOL model.

| Parameter | Symbol | Value |
| --- | --- | --- |
| Density of Pyromark coating | $\rho_{Pyromark}$ | 1100 kg m$^{-3}$ |
| Density of stainless-steel sheet | $\rho_{SS}$ | 7751 kg m$^{-3}$ |
| Density of air | $\rho_{air}$ | 0.616 (300 °C), 0.488 (450 °C), 0.383 (650 °C) kg m$^{-3}$ |
| Density of particle bed | $\rho_{bed}$ | 2278 kg m$^{-3}$ |
| Specific heat of Pyromark coating | $c_{Pyromark}$ | 1100 (300 °C), 1471 (450 °C), 1558 (650 °C) J kg$^{-1}$ K$^{-1}$ |
| Specific heat of stainless-steel sheet | $c_{SS}$ | 535 (300 °C), 555 (450 °C), 582 (650 °C) J kg$^{-1}$ K$^{-1}$ |
| Specific heat of air | $c_{air}$ | 1044 (300 °C), 1081 (450 °C), 1125 (650 °C) J kg$^{-1}$ K$^{-1}$ |

| Specific heat of particle bed | $c_{bed}$ | 1038 (300 °C), 1101 (450 °C), 1130 (650 °C) J kg$^{-1}$ K$^{-1}$ |
|---|---|---|
| Thermal conductivity of Pyromark coating | $k_{Pyromark}$ | 0.5 W m$^{-1}$ K$^{-1}$ |
| Thermal conductivity of stainless-steel sheet | $k_{SS}$ | 18.3 (300 °C), 20.6 (450 °C), 23.8 (650 °C) W m$^{-1}$ K$^{-1}$ |
| Thermal conductivity of air | $k_{air}$ | 0.044 (300 °C), 0.053 (450 °C), 0.063 (650 °C) W m$^{-1}$ K$^{-1}$ |
| Laser heat flux | $q_0$ | 5118.57 W m$^{-2}$ |
| Laser beam width | $D_{laser}$ | 20 mm |
| Thickness of Pyromark coating | $D_{Pyromark}$ | 10 μm |
| Thickness of stainless-steel sheet | $D_{SS}$ | 100 μm |
| Channel depth | $D_{channel}$ | 5 mm |

Notes:

a. The density, specific heat, and thermal conductivity of the Pyromark coating were measured at various temperatures in our previous work using MPR to characterize bulk materials [1].

b. The density, specific heat, and thermal conductivity of the stainless steel 316 sheet were based on measurement results from Argonne National Laboratory [4].

c. The density, specific heat, and thermal conductivity of air were from the air property table in the textbook [5]

d. The effective density of ceramic particle bed was based on the measurement mentioned in this Supplemental Material, section S2. The specific heat of ceramic particles was measured by differential scanning calorimetry with a Netzsch STA 409 [6].

e. The laser heat flux was calibrated following the procedures introduced in Supplemental Material, S3. The laser beam width was measured with a scale during experiments.

f. The thickness of the Pyromark coating was confirmed by scanning electron microscopy (SEM) imaging shown in Fig. S10.

g. The thickness of the stainless-steel sheet was claimed to be 0.004 inch-thick (100 μm-thick) by the vendor [7] and was confirmed by a micrometer.

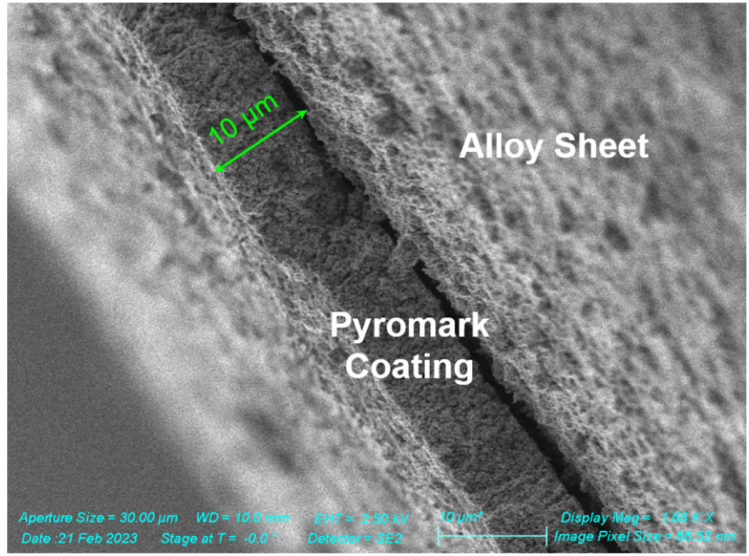

FIG. S10. SEM image of Pyromark coated on alloy sheet.

**S7. Translational and rotational kinetic energy of particles from DEM**

We examined the average translational kinetic energy $\bar{E}_{k,t}$ and the average rotational kinetic energy $\bar{E}_{k,r}$ of flowing HSP 40/70 particles in the *x-y* plane shown in Fig. 3(a). Fig. S11 shows that the $\bar{E}_{k,r}$ is around two times larger than the $\bar{E}_{k,t}$ in the first layer of particles adjacent to the wall (at $y = 0.5D_p$) due to the wall friction. From $y = 0.5D_p$ to $y = 1.0D_p$, $\bar{E}_{k,r}$ decreases sharply and $\bar{E}_{k,t}$ increases to its maximum, implying the conversion of kinetic energy during the interaction between the first and the second layers of particles. From $y = 1.0D_p$ to the channel center, both $\bar{E}_{k,t}$ and $\bar{E}_{k,r}$ monotonically decrease, which represents the dissipation of shear work introduced by the wall. Here, we only show the results of HSP 40/70 but the conclusion holds for CP 40/100 as well.

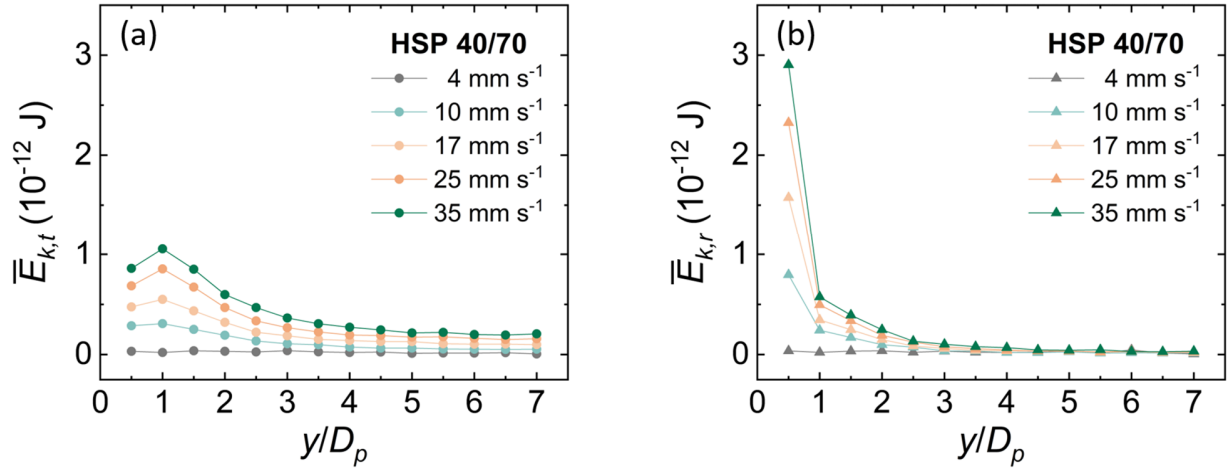

FIG. S11. (a) Average translational kinetic energy $\bar{E}_{k,t}$ and (b) average rotational kinetic energy $\bar{E}_{k,r}$ of flowing HSP 40/70 along the *y* direction from the DEM simulation.

**References**

[1] J. Zeng, K. M. Chung, Q. Wang, X. Wang, Y. Pei, P. Li, and R. Chen, *International Journal of Heat and Mass Transfer* **170**, 120989 (2021).

[2] K. M. Chung, J. Zeng, S. R. Adapa, T. Feng, M. V. Bagepalli, P. G. Loutzenhiser, K. J. Albrecht, C. K. Ho, and R. Chen, Solar Energy Materials and Solar Cells **230**, 111271 (2021).

[3] B. Cheng, B. Lane, J. Whiting, and K. Chou, Journal of manufacturing science and engineering **140**, 111008 (2018).

[4] Kim, C.S., 1975. *Thermophysical properties of stainless steels* (No. ANL-75-55). Argonne National Lab., Ill.(USA).

[5] Yunus, A.C., 2010. *Fluid Mechanics: Fundamentals And Applications (Si Units)*. Tata McGraw Hill Education Private Limited.

[6] Siegel, N.P., Gross, M.D. and Coury, R. *Journal of Solar Energy Engineering* **137**, p.041003 (2015).

[7] https://www.mcmaster.com/products/sheets/shim-stock-6/